\documentclass[conference]{IEEEtran}
\IEEEoverridecommandlockouts

\usepackage[utf8]{inputenc}
\usepackage[T1]{fontenc}
\usepackage{graphicx}
\usepackage{amsmath}
\usepackage{amsfonts}
\usepackage{amssymb}
\usepackage{booktabs}
\usepackage{multirow}
\usepackage{siunitx}
\usepackage{cite}
\usepackage{url}

\title{A Proposal for Yield Improvement with Power Tradeoffs in CMOS LNAs (English Version)}

\author{\IEEEauthorblockN{J. L. Gonz\'alez, J. C. Cruz, R. L. Moreno, D. V\'azquez}}

\begin{document}
\maketitle

\begin{abstract}
This paper studies an architecture with digitally controllable gain and power consumption to mitigate the impact of process variations on CMOS low-noise amplifiers (LNAs). A \SI{130}{nm}, \SI{1.2}{V} LNA implementing the proposed architecture is designed based on an analysis of variability in traditional LNAs under different bias currents and on the corresponding effects on the performance of a complete receiver. Two different adjustment strategies are evaluated, both of which are compatible with previously reported built-in self-test (BIST) circuits. Results show that the proposed architecture enables yield enhancement while keeping low-power operation compared with traditional LNAs.
\end{abstract}


\section{Introduction}
Short-range wireless communication devices are continually expanding their application scope, which poses several design challenges, including miniaturization, low-voltage operation, and low power consumption. Data transfer in these applications also requires receivers with a wide dynamic range due to the variability of signal levels in the radio-frequency (RF) channel and the presence of high interference~\cite{tang2007}. 
Fulfilling these demands introduces several design trade-offs in the implementation of the different blocks of the RF front-end~\cite{razavi1998}, primarily in the low-noise amplifier (LNA), the first active block of the receiver.

The LNA determines the receiver's sensitivity through sufficiently high gain and a low noise figure~\cite{razavi1998,lee2004}. 
However, excessive gain may saturate subsequent blocks (e.g., the mixer) under high input levels, posing a design trade-off. 
At the same time, the LNA must provide good input impedance matching, sufficiently high linearity, and enough reverse isolation.

\begin{figure}[!t]
    \centering
    \includegraphics[width=\columnwidth]{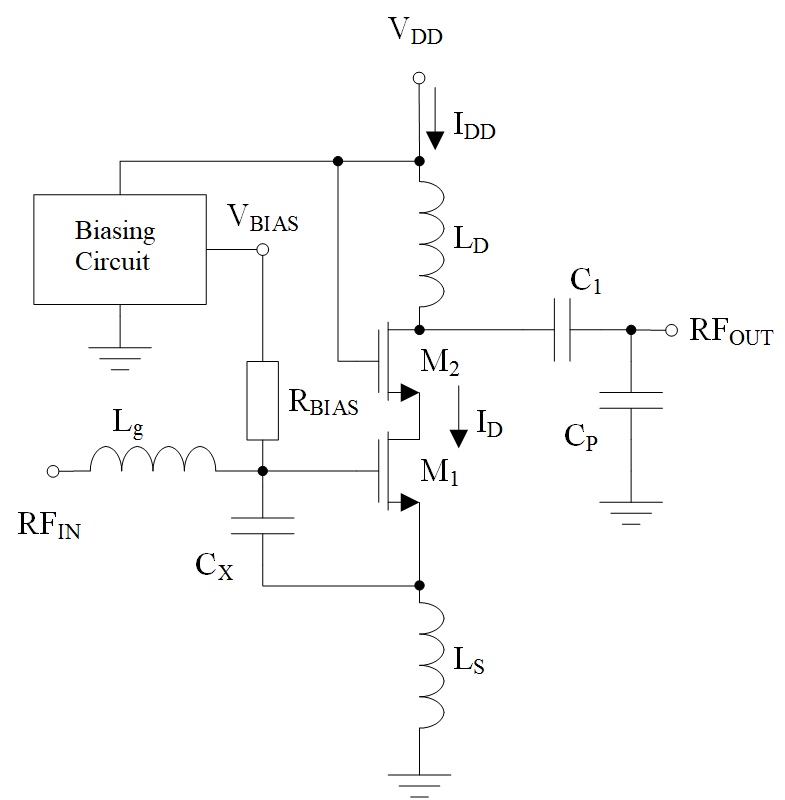}
    \caption{Common-source LNA with inductive source degeneration.}
    \label{fig1}
\end{figure}

The common-source LNA with inductive source degeneration (Fig.~\ref{fig1}) is a widely used topology in integrated CMOS receivers for short-range communication systems~\cite {lee2004,leroux2005,farahani2008}. 
Design methodologies for this topology are typically focused on noise figure and power minimization~\cite{shaeffer1997,belostotski2006,fiorelli2014}. 
Moreover, the design flow must account for the effects of process variations, given their significant impact in deep-submicron CMOS technologies~\cite{kuhn2011,alam2011}. 
Several works have analyzed the impact of process variations on this topology and ways to mitigate their consequences~\cite{sivonen2004,nieuwoudt2008,sen2008,gomez2010,yidong2011,mukadam2014}. 
Nevertheless, those studies focus on specific LNA parameters without analyzing their relationships to receiver-level performance, nor the influence of power consumption on the effects of process variations.

In this work, we employ an architecture with digitally controllable gain and power consumption proposed in 
\cite{gonzalez2013lna} 
to mitigate process-variation effects in a \SI{130}{nm}, \SI{1.2}{V} CMOS LNA. 
The design is grounded on Monte Carlo simulations that analyze parameter variability in the traditional LNA and its influence on receiver performance across different bias current levels.
Two adjustment strategies are evaluated: (i) a method based on measuring the deviation of gain with respect to its nominal value and (ii) a method based on receiver-level performance parameters. Results show that the proposed architecture reduces the number of receiver failures caused by process variations in the LNA while preserving low power consumption compared to the conventional architecture.

\section{Analysis and Design of Traditional LNAs}
\subsection{Topology}
Figure~\ref{fig1} shows the basic schematic of a common-source (CS) LNA with inductive degeneration. Inductive degeneration (through $L_s$) produces a resistive component in the input impedance without introducing an extra noise source~\cite{karanicolas1996}. Capacitor $C_X$ helps minimize the noise figure (NF) for specific gain and power levels~\cite{andreani2001}. The gate inductor $L_g$ is included to tune the input impedance. Transistor $M_2$ is used as a cascode stage to reduce the Miller effect on $M_1$ and improve reverse isolation~\cite{lee2004}. The drain inductor $L_D$ forms a parallel resonant network together with the output capacitances of the cascode and the load impedance. Finally, the capacitive divider ($C_1$, $C_P$) is included to match the output impedance to \SI{50}{\ohm} for characterization of the standalone LNA with a spectrum analyzer.

In the design of a CS-LNA for a given gain and power, the transistors can be sized to minimize the noise figure, as shown in prior work for this topology~\cite{shaeffer1997,belostotski2006,andreani2001}. High input-referred third-order intercept point (IIP3) at low power can be achieved by biasing MOS transistors in moderate inversion to exploit the linearity sweet spot~\cite{aparin2004}. This IIP3 peak occurs approximately at the same current density in the common-source device~\cite{niu2005,toole2004}. Therefore, for a given power budget, linearity can also be maximized by adjusting the device size appropriately.

Based on this topology, we designed several LNAs with different bias currents. Then we studied the effects of process variations on their parameters and the expected impact on overall receiver performance.

\subsection{Design Space Exploration}
The LNAs were sized by exploring the design space in the available technology, targeting the RF specifications in Table~\ref{tab:specs} for a ZigBee/IEEE~802.15.4 receiver~\cite{trung-kien2006,fiorelli2011}. Input and output reflection coefficients ($S_{11}$ and $S_{22}$) are referred to \SI{50}{\ohm}.

\begin{table}[!t]
    \centering
    \caption{LNA specifications (matching at \SI{50}{\ohm}).}
    \label{tab:specs}
    \begin{tabular}{@{}ll@{}}
        \toprule
        Frequency & \SI{2.4}{}--\SI{2.5}{GHz} \\
        Gain $G$ & $10.5 \pm 0.5$~dB \\
        Noise figure (NF) & $< 3$~dB \\
        IIP3 & $> -4$~dBm \\
        $S_{11}$, $S_{22}$ & $< -10$~dB \\
        \bottomrule
    \end{tabular}
\end{table}

We swept the bias current $I_D$ and the width of the transistor $M_1$ ($W_1$). The width of $M_2$ was set to $W_2 = W_1/2$ to reduce its contribution to load capacitance and enlarge the selection margin for the output matching network~\cite{leroux2005}. The channel length of all transistors was fixed to the minimum value allowed by the technology, $L_1=L_2=L=\SI{0.12}{\micro m}$. For each combination of current and transistor widths, the passive components were chosen to meet the gain and matching requirements.

\begin{figure}[!t]
    \centering
    \includegraphics[width=\columnwidth]{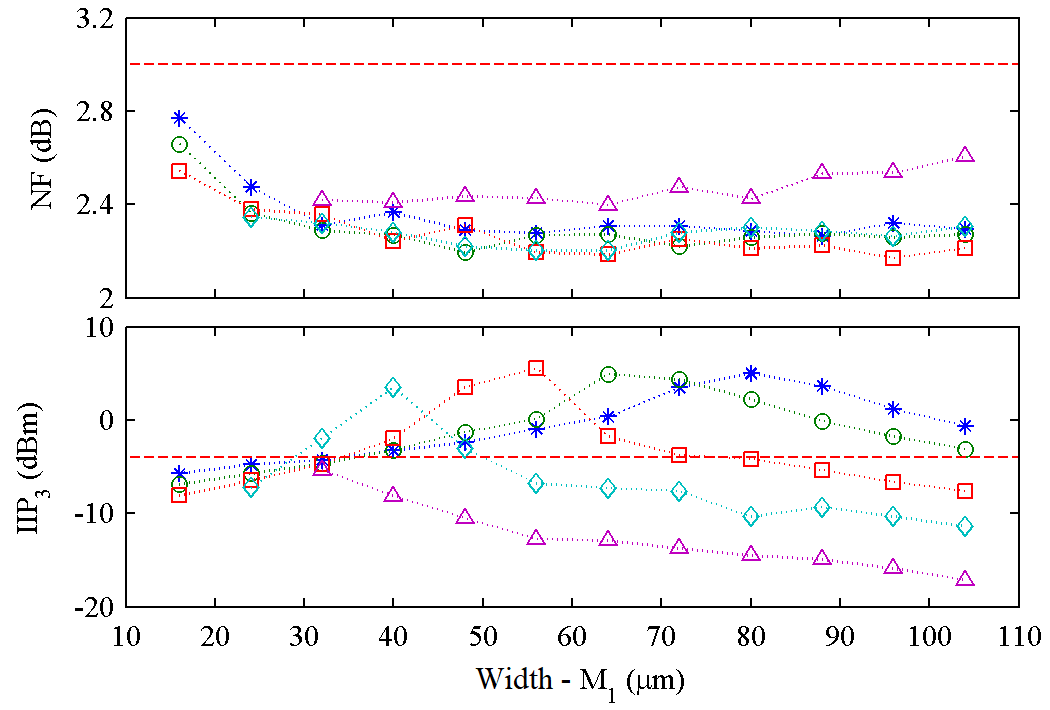}
    \caption{Simulations at \SI{2.45}{GHz} for NF (top) and IIP3 (bottom) versus $W_1$ and $I_D$ ($\triangle$: \SI{0.3}{mA}, $\diamond$: \SI{0.4}{mA}, $\square$: \SI{0.5}{mA}, $\circ$: \SI{0.6}{mA}, $*$:\SI{0.7}{mA}). Horizontal dashed lines mark the requirements.}
    \label{fig2}
\end{figure}

Figure~\ref{fig2} shows simulated results at \SI{2.45}{GHz} for NF and IIP3. All the shown designs meet $S_{11}, S_{22}<-15$~dB and gain in $[10.3, 10.9]$~dB. Minimum values of $W_1$ for $I_D=\SI{0.4}{mA}$ (\SI{24}{\micro m}) and $I_D=\SI{0.3}{mA}$ (\SI{32}{\micro m}) were limited by the practical values of the passive elements in the technology while still satisfying gain and matching. All synthesized LNAs met the NF specification (NF~$<3$~dB). However, the linearity requirement (IIP3~$>-4$~dBm) is not met for $I_D=\SI{0.3}{mA}$, which is therefore discarded as a bias option, and no lower currents were consequently analyzed.

As case studies, we selected amplifiers with $I_D \in \{0.4, 0.5, 0.6, 0.7\}$~mA. For each power level, we chose the LNA with the highest IIP3. The resulting sizing is listed in Table~\ref{tab:dim_trad}. A simple current mirror was used as the biasing circuit.

\begin{table}[!t]
\centering
\caption{Sizing of the selected LNAs.}
\label{tab:dim_trad}
\begin{tabular}{@{}lcccc@{}}
\toprule
\multirow{2}{*}{Parameter} & \multicolumn{4}{c}{$I_{NOM}$ (mA)} \\ \cmidrule(l){2-5}
 & 0.4 & 0.5 & 0.6 & 0.7 \\
\midrule
$W_1/L_1$ ($\mu$m/$\mu$m) & 40/0.12 & 56/0.12 & 64/0.12 & 80/0.12 \\
$L_s$ (nH) & 2.51 & 2.51 & 2.51 & 2.65 \\
$L_g$ (nH) & 11.8 & 7.40 & 6.06 & 5.02 \\
$C_X$ (fF) & 246 & 383 & 453 & 532 \\
$C_1$ (fF) & 439 & 429 & 425 & 416 \\
$C_P$ (pF) & 1.71 & 1.62 & 1.61 & 1.54 \\
$L_D$ (nH) & 10.5 & 10.5 & 10.5 & 10.5 \\
\bottomrule
\end{tabular}
\end{table}

\subsection{Variability Analysis}
For each selected LNA, we performed a Monte Carlo simulation with 1000 runs, incorporating both process and mismatch variations, as specified by the PDK's statistical distributions. Table~\ref{tab:mc_trad_values} summarizes mean and worst-case values of the simulated parameters, while Table~\ref{tab:mc_trad_viol} lists the percentage of specification violations.

\begin{table*}[!t]
\centering
\caption{Mean and worst-case values from Monte Carlo (traditional LNAs).}
\label{tab:mc_trad_values}
\begin{tabular}{@{}lcccccccccccc@{}}
\toprule
 & \multicolumn{3}{c}{$I_{NOM}=\SI{0.4}{mA}$} & \multicolumn{3}{c}{$I_{NOM}=\SI{0.5}{mA}$} & \multicolumn{3}{c}{$I_{NOM}=\SI{0.6}{mA}$} & \multicolumn{3}{c}{$I_{NOM}=\SI{0.7}{mA}$} \\ \cmidrule(l){2-13}
Parameter & min & mean & max & min & mean & max & min & mean & max & min & mean & max \\
\midrule
$G$ (dB)   & 8.64 & 10.4 & 11.7 & 8.23 & 10.4 & 11.6 & 8.92 & 10.5 & 11.6 & 8.67 & 10.4 & 11.4 \\
NF (dB)    &  --  &  2.8 &  --  &  --  &  2.8 &  --  &  --  &  2.7 &  --  &  --  &  2.7 &  --  \\
IIP3 (dBm) & -10.4 & 0.7 &  --  & -9.6 & 4.2 &  --  & -6.6 & 5.4 &  --  & -5.8 & 5.4 &  --  \\
$S_{11}$ (dB) &  -- & -15 & -- & -- & -16 & -- & -- & -17 & -- & -- & -17 & -- \\
$S_{22}$ (dB) &  -- & -6.1 & -- & -- & -6.3 & -- & -- & -6.4 & -- & -- & -6.8 & -- \\
\bottomrule
\end{tabular}
\end{table*}

\begin{table}[!t]
\centering
\caption{Spec-violation percentages (traditional LNAs).}
\label{tab:mc_trad_viol}
\begin{tabular}{@{}lcccc@{}}
\toprule
 & \multicolumn{4}{c}{$I_{NOM}$ (mA)} \\ \cmidrule(l){2-5}
Violation & 0.4 & 0.5 & 0.6 & 0.7 \\
\midrule
$G<10$~dB & 22\% & 23\% & 16\% & 16\% \\
$G>11$~dB & 11\% & 10\% & 11\% & 7\% \\
NF $>3$~dB & 0\% & 0\% & 0\% & 0\% \\
IIP3 $<-4$~dBm & 14\% & 8\% & 1\% & 0\% (1/1000) \\
$S_{11}>-10$~dB & 0\% & 0\% & 0\% & 0\% \\
$S_{22}>-10$~dB & 3\% & 3\% & 2\% & 2\% \\
\bottomrule
\end{tabular}
\end{table}

From Table~\ref{tab:mc_trad_values}, the mean IIP3 increases as bias current goes from \SI{0.4}{mA} to \SI{0.6}{mA}. The mean values of the remaining parameters vary little across designs. In worst-case terms, IIP3 exhibits the largest improvement as the current increases. Input matching ($S_{11}$) and noise figure specifications are met in all cases, whereas output matching ($S_{22}$) shows a low probability of violation ($<3\%$). The number of IIP3 violations drops from 14\% to just 1\% when increasing the current from \SI{0.4}{mA} to \SI{0.6}{mA}.

By contrast, gain is the parameter with the highest probability of falling outside the specification, most often due to $G < 10$~dB. This is significant because gain variations affect both receiver noise and linearity, even when NF and IIP3 of the LNA itself meet their limits.

The implications at receiver level can be studied using the cascade equations that relate LNA parameters ($F_{\mathrm{LNA}}$, $\mathrm{IIP3}_{\mathrm{LNA}}$) to receiver parameters ($F_{\mathrm{Rx}}$, $\mathrm{IIP3}_{\mathrm{Rx}}$) and to the subsequent stages ($F_2$, $\mathrm{IIP3}_2$)~\cite{razavi1998}:
\begin{align}
F_{\mathrm{Rx}} &= F_{\mathrm{LNA}} + \frac{F_2 - 1}{G_{\mathrm{LNA}}}, \label{eq:friis}\\
\frac{1}{\mathrm{IIP3}_{\mathrm{Rx}}} &= \frac{1}{\mathrm{IIP3}_{\mathrm{LNA}}} + \frac{G_{\mathrm{LNA}}}{\mathrm{IIP3}_2}. \label{eq:iip3cascade}
\end{align}
where $F$ denotes the noise factor ($F=10^{\mathrm{NF}/10}$) and $G_{\mathrm{LNA}}$ is the linear power gain.

Using the LNA specs of Table~\ref{tab:specs} and receiver targets ($\mathrm{NF}_{\mathrm{Rx}}\leq \SI{15.5}{dB}$ and $\mathrm{IIP3}_{\mathrm{Rx}}\geq \SI{-10}{dBm}$)~\cite{trung-kien2006}, the limits for the remainder of the chain can be computed:
\begin{align}
F_2 &\le F_{2,\mathrm{max}}, \qquad
\mathrm{IIP3}_2 \ge \mathrm{IIP3}_{2,\mathrm{min}}.
\end{align}
Assuming receivers that include the simulated LNAs and subsequent stages sized at these limits, such receivers meet the specs if
\begin{align}
F_{\mathrm{Rx}} &\le F_{\mathrm{Rx,max}}, \quad
\mathrm{IIP3}_{\mathrm{Rx}} \ge \mathrm{IIP3}_{\mathrm{Rx,min}}.
\end{align}
Table~\ref{tab:rx_stats_trad} summarizes the percentage of receivers that simultaneously meet both noise and linearity specs and those failing noise or linearity, respectively.

\begin{table}[!t]
\centering
\caption{Receiver-level compliance based on traditional LNAs.}
\label{tab:rx_stats_trad}
\begin{tabular}{@{}lcccc@{}}
\toprule
 & \multicolumn{4}{c}{$I_{NOM}$ (mA)} \\ \cmidrule(l){2-5}
Outcome & 0.4 & 0.5 & 0.6 & 0.7 \\
\midrule
Meet both specs & 77\% & 79\% & 86\% & 86\% \\
NF out of spec & 21\% & 21\% & 14\% & 14\% \\
IIP3 out of spec & 6\% & 0\% (3/1000) & 0\% & 0\% \\
\bottomrule
\end{tabular}
\end{table}

The LNA parameter variations affect receiver noise more than linearity. When $G$ drops below its minimum, the LNA cannot suppress the noise of the following stages even if $F_{\mathrm{LNA}}$ itself is small. In contrast, receiver linearity is less sensitive to the LNA since this is the stage with the lowest input signal level; thus, maintaining a high $\mathrm{IIP3}_{\mathrm{LNA}}$ prevents most linearity failures even when gain increases.

One possible solution to compensate for gain loss---and thus increase the number of compliant cases---would be to improve the noise figure of the subsequent stages, primarily the mixer. However, this would complicate implementation because system-level specifications would need to be redistributed, and some blocks would require redesign. Moreover, reducing the noise figure typically requires higher power consumption in the corresponding block~\cite{sheng2006}. 
To avoid these drawbacks, we evaluate an LNA architecture with controllable gain and power to mitigate process-variation effects at the LNA itself.

\section{Using a Programmable Architecture to Mitigate LNA Performance Degradation}
\begin{figure}[!t]
    \centering
    \includegraphics[width=\columnwidth]{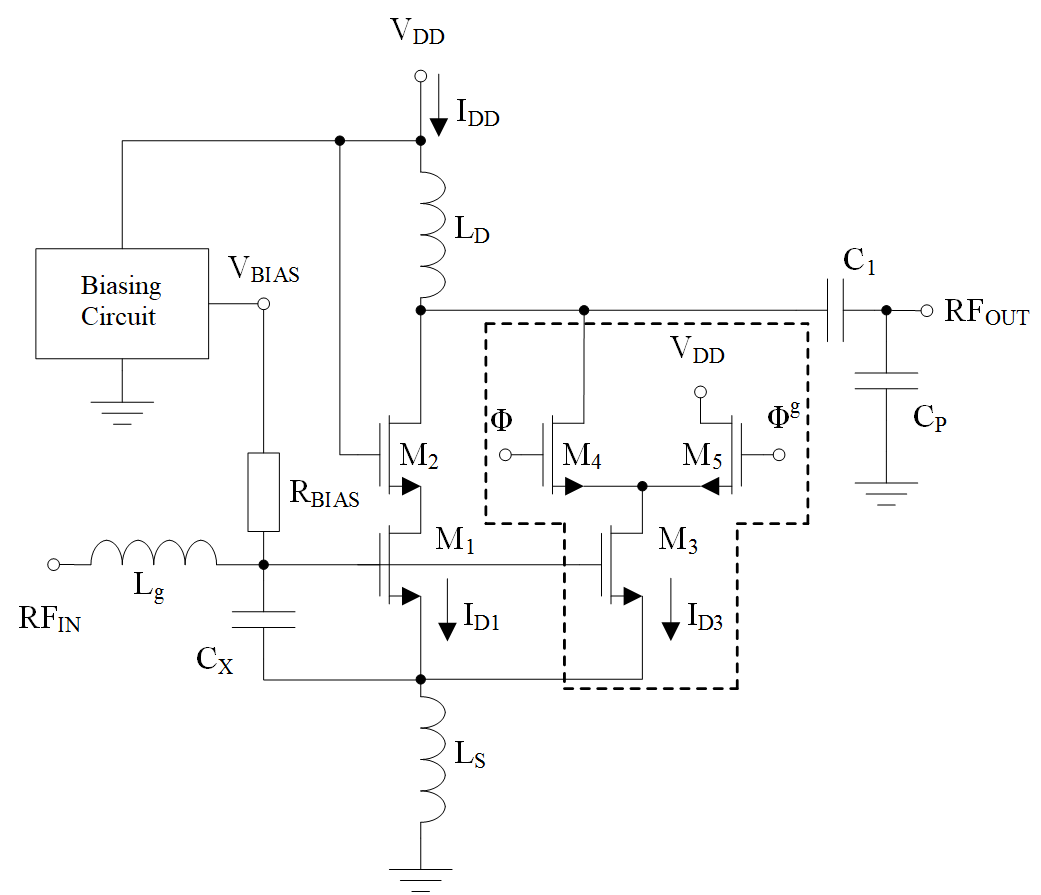}
    \caption{Programmable architecture to mitigate parameter degradation in the LNA.}
    \label{fig3}
\end{figure}

The programmable LNA (PLNA) architecture proposed in 
\cite{gonzalez2013lna} and shown in Fig.~\ref{fig3} 
extends the traditional CS-LNA. An extra parallel branch is added, comprised of a transconductor ($M_3$) and a cascode pair that acts as a current switch ($M_4$ to the output network, $M_5$ to $V_{DD}$). This branch combines two control techniques: gain control via a current-splitting variant~\cite{koutani2007} and power control via transistor width scaling~\cite{ghosal2012}. The topology supports several digitally selected operating modes through the control inputs $\Phi$ and $\Phi_g$, providing both gain and power control.

Complementary switching $M_4$ and $M_5$ ($\Phi=1/0$, $\Phi_g=0/1$) either delivers the RF current flowing through $M_3$ to the load (high-gain mode, \textbf{HG}) or diverts it to ground via $V_{DD}$ (low-gain mode, \textbf{LG}). Because the operating point of $M_3$ remains constant during switching, the input impedance is unaffected.

When both $M_4$ and $M_5$ are off ($\Phi=\Phi_g=0$), current through $M_3$ ceases, which reduces power but also changes the input impedance. In this mode, the RF signal at the output flows only through $M_2$, as in LG. However, the change in input impedance makes the gain values differ; we observed that the lowest-power mode provides an intermediate gain. Therefore, the combination $\Phi=\Phi_g=0$ yields a medium-gain, low-power mode (\textbf{MG-LP}).

In the implemented power scaling, with fixed gate bias for the common-source devices, the equivalent current density is kept constant,
\begin{equation}
\frac{I_{D1}}{W_1}=\frac{I_{D1}+I_{D3}}{W_1+W_3},
\end{equation}
which allows all modes to operate around the current-density region where the IIP3 peak appears. From the higher-current modes, one also expects more stable linearity under process variations, consistent with the traditional-LNA results.

The ratio $W_1/W_3$ determines both the achievable gain steps and the change in bias current. This ratio is limited by the permissible degradation in input matching, which sets a design tradeoff for the PLNA \cite{gonzalez2016clna}.

\subsection{Design}
We set the medium gain equal to the previous target of \SI{10.5}{dB}. Simulations showed that a $3{:}1$ ratio between $W_1$ and $W_3$ provides $+1.5$~dB and $-1.0$~dB gain steps around the medium-gain value while maintaining acceptable input matching across modes. Hence, a worst-case $-2$~dB gain drop (i.e., \SI{8.5}{dB} in MG-LP, similar to the worst cases in Table~\ref{tab:mc_trad_values}) can be reduced to a $-0.5$~dB deviation by switching to HG. Likewise, LG can recover worst-case scenarios with a $+1.5$~dB excess gain back to within specification.

With this aspect ratio, $M_1$ and $M_3$ were dimensioned and biased near the IIP3 sweet spot (Fig.~\ref{fig2}) while minimizing power. We selected $\{W_1;I_{D1}\}=\{3W_3;3I_{D3}\}=\{\SI{42}{\micro m};\SI{0.4}{mA}\}$ for MG-LP, which yields the equivalent operating point $\{W_1+W_3; I_{D1}+I_{D3}\}=\{4W_3;4I_{D3}\}=\{\SI{56}{\micro m};\SI{0.53}{mA}\}$ for the higher-power modes (HG and LG). The final dimensions are listed in Table~\ref{tab:dim_plna}. The value of $L_D$ is the same as in Table~\ref{tab:dim_trad}. $I_{DD}$ includes the current delivered to the bias circuitry.

\begin{table}[!t]
\centering
\caption{Dimensions of the programmable LNA (PLNA).}
\label{tab:dim_plna}
\begin{tabular}{@{}lcc@{}}
\toprule
Parameter & MG-LP / HG,LG \\
\midrule
$I_{DD}$ (mA) & 0.43 / 0.56 \\
$W_1/L_1$ ($\mu$m/$\mu$m) & 42/0.12 \\
$W_3/L_3$ ($\mu$m/$\mu$m) & 14/0.12 \\
$L_s$ (nH) & 2.38 \\
$L_g$ (nH) & 11.23 \\
$C_X$ (fF) & 246 \\
$C_1$ (fF) & 426 \\
$C_P$ (pF) & 1.59 \\
$L_D$ (nH) & 10.5 \\
\bottomrule
\end{tabular}
\end{table}

The passive components of the PLNA differ only slightly from those of the \SI{0.4}{mA} traditional LNA ($<5\%$ in inductors and $<7\%$ in capacitors). Therefore, the occupied die area (mainly determined by the inductor sizes) is expected to be practically unaffected by the programmable architecture.

\subsection{Variability Analysis}
We ran 1000-run Monte Carlo simulations for each PLNA mode. Table~\ref{tab:plna_mc} lists mean and worst-case values.

\begin{table*}[!t]
\centering
\caption{Mean and worst-case values from Monte Carlo (PLNA modes).}
\label{tab:plna_mc}
\begin{tabular}{@{}lccccccccc@{}}
\toprule
 & \multicolumn{3}{c}{HG} & \multicolumn{3}{c}{MG-LP} & \multicolumn{3}{c}{LG} \\ \cmidrule(l){2-10}
Parameter & min & mean & max & min & mean & max & min & mean & max \\
\midrule
$G$ (dB)   & 10.1 & 11.9 & 13.1 & 8.48 & 10.5 & 11.9 & 7.58 & 9.5 & 10.8 \\
NF (dB)    &  --  &  2.5 &  --  &  --  &  2.9 &  --  &  --  &  3.6 &  --  \\
IIP3 (dBm) & -8.9 & 2.3  &  --  & -11.0 & -1.5 &  --  & -9.2 & 2.4 &  --  \\
$S_{11}$ (dB) &  -- & -14 & -- & -- & -16 & -- & -- & -14 & -- \\
$S_{22}$ (dB) &  -- & -8.3 & -- & -- & -8.1 & -- & -- & -8.2 & -- \\
\bottomrule
\end{tabular}
\end{table*}

Comparing MG-LP to the \SI{0.4}{mA} traditional LNA, only IIP3 shows a lower mean value, yet it still meets the spec. Worst-case behavior is similar in both amplifiers. Therefore, introducing programmability does not practically degrade the base amplifier performance. In the higher-power modes, IIP3 improves while input matching degrades (yet remains within limits). The minimum and maximum gains in HG and LG, respectively, also comply with the required range.

\section{PLNA Results With Mode Selection Under Process Variations}
From the Monte Carlo results, for each sample, we selected the operating mode according to two criteria:

\begin{itemize}
\item \textbf{Best Gain}: choose the mode with the smallest deviation of LNA gain from \SI{10.5}{dB}.
\item \textbf{Best Receiver}: choose the mode for which the receiver meets both NF and linearity limits in~\eqref{eq:friis}--\eqref{eq:iip3cascade}. If this is not possible, select a mode that at least meets the NF condition; if none is possible, choose the mode with the largest dynamic range.
\end{itemize}

These two criteria were considered because both are compatible with integrated BIST solutions~\cite{yen-chih2008,senguttuvan2008,barragan2010} that could enable automatic adjustment.

Table~\ref{tab:plna_sel_values} shows mean and worst-case RF parameters after mode selection for each criterion, and Table~\ref{tab:plna_sel_rx} summarizes receiver-level compliance and failures.

\begin{table}[!t]
\centering
\caption{Mean and worst-case values after PLNA mode selection.}
\label{tab:plna_sel_values}
\begin{tabular}{ccccccc}
\hline
\multirow{2}{*}{Parameter} & \multicolumn{3}{c}{Best Gain} & \multicolumn{3}{c}{Best Receiver} \\ \cline{2-7} 
                           & min      & mean     & max     & min       & mean      & max       \\ \hline
G (dB)                     & 9.73     & 10.5     & 11.3    & 9.97      & 10.7      & 11.8      \\ \hline
NF (dB)                    &          & 2.4      & 2.99    &           & 2.3       & 2.98      \\ \hline
IIP3 (dBm)                 & -9.0     & -0.7     &         & -8.9      & -0.9      &           \\ \hline
S11 (dB)                   &          & -23      & -14     &           & -24       & -14       \\ \hline
S22 (dB)                   &          & -19      & -8.1    &           & -19       & -8.1      \\ \hline
\end{tabular}
\end{table}

\begin{table}[!t]
\centering
\caption{Receiver-level compliance after PLNA mode selection.}
\label{tab:plna_sel_rx}
\begin{tabular}{@{}lcc@{}}
\toprule
Outcome & Best Gain & Best Receiver \\
\midrule
Meet both specs & 85\% & 92\% \\
NF out of spec & 7\% & 0\% \\
IIP3 out of spec & 9\% & 8\% \\
\bottomrule
\end{tabular}
\end{table}

Both selection methods yield very similar worst-case scenarios, except for gain, which is higher when prioritizing receiver compliance because this criterion favors NF improvement; as a side effect, the maximum gain increases. As expected, the receiver-based selection increases the fraction of fully compliant receivers. With this method, we fully compensated all cases where gain loss caused the receiver NF to fail. The two methods show approximately the same number of cases where linearity could not be compensated. Selecting one method or the other will also depend on the practical pros and cons of the corresponding BIST implementations.

\subsection{Performance Comparison With Traditional LNAs}
Compared to the traditional LNAs, the PLNA with either selection method shows clear improvements. In worst-case terms, only NF and $S_{11}$ degrade after adjustment. NF degradation occurs due to the LG mode~\cite{kengleong1999}, while $S_{11}$ degradation stems from operating-point changes; yet all cases remain within limits.

\begin{figure}[!t]
    \centering
    \includegraphics[width=\columnwidth]{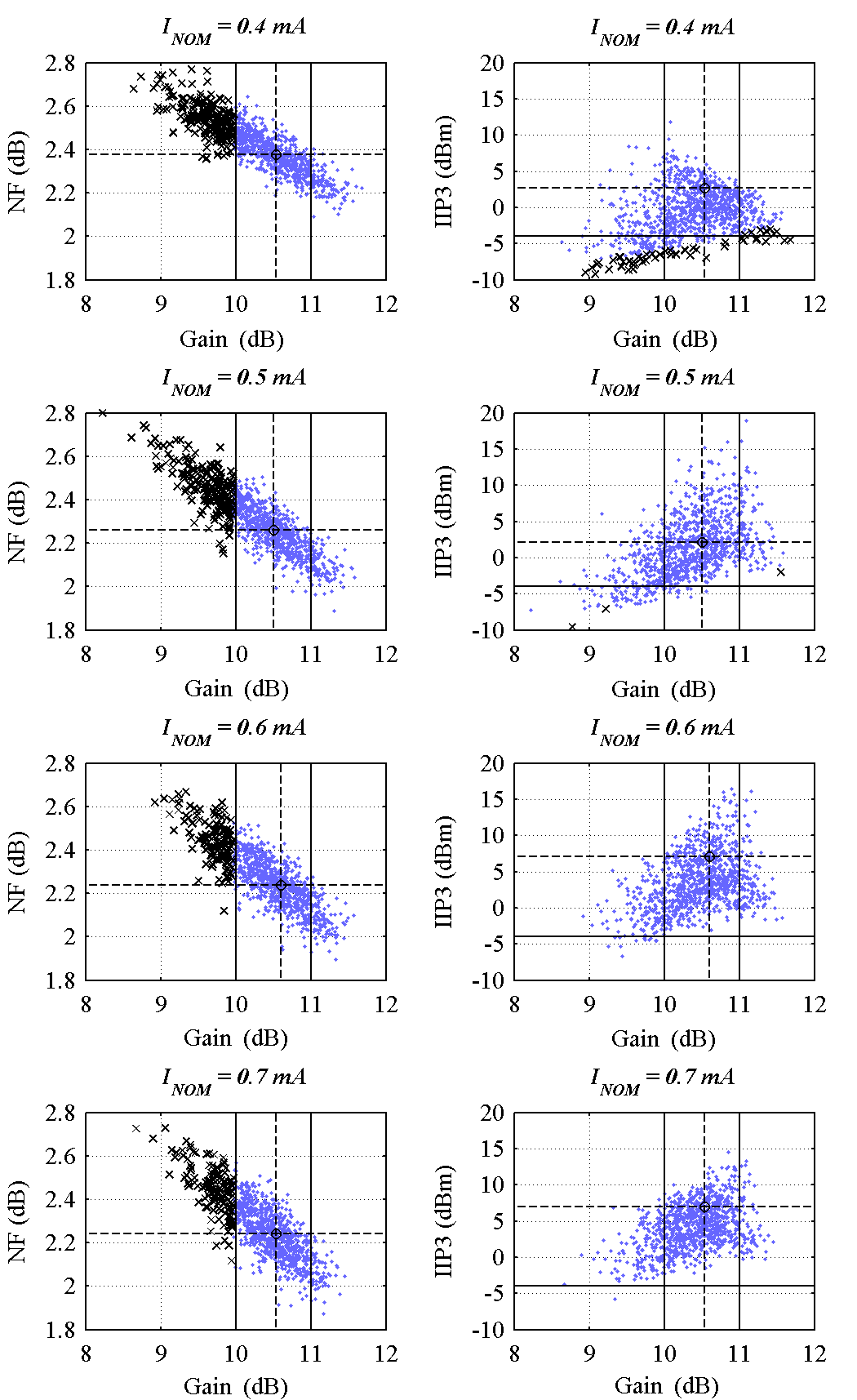}
    \caption{Comparison of PLNA (top: Best Gain selection; bottom: Best Receiver selection) versus traditional LNAs in terms of differences in the number of compliant cases and in average power consumption.}
    \label{fig4}
\end{figure}

Figure~\ref{fig4} compares the PLNA performance after adjustment (both selection methods) with the traditional LNAs in terms of the difference in the number of compliant cases ($\Delta S$) and the average power consumption difference ($\Delta P$), each normalized to the total number of cases and to the power of each LNA, respectively. The ``Best Gain'' method reaches practically the same compliance rate as the \SI{0.6}{mA} and \SI{0.7}{mA} LNAs (differences $<1\%$) with 26\% and 35\% lower average power, respectively. It also improves the compliance rate of the \SI{0.4}{mA} LNA by 8\% at the cost of only 9\% more power---a balanced tradeoff. The ``Best Receiver'' method exceeds the compliance rate of all analyzed traditional LNAs with average power only higher than that of the \SI{0.4}{mA} design; relative to that design, it consumes merely 5\% more power while delivering a 15\% improvement in compliance.

\section{Conclusions}
We presented a digitally controllable-gain, power-scalable LNA architecture to mitigate process-variation effects in a \SI{130}{nm}, \SI{1.2}{V} CMOS technology. By adjusting transistor operating points through digital inputs as a function of RF parameter degradation, the architecture reduces the number of receiver failures caused by LNA variability while maintaining low power consumption compared to traditional LNA designs.

Two different adjustment strategies were evaluated: one based on gain deviation and another based on receiver performance, both of which outperformed the traditional LNA. The second method shows a better balance between the number of compliant receivers and average power consumption.

The proposed architecture requires adding only a few devices (transistors) to the standard common-source, inductively degenerated LNA topology. Passive components change only slightly compared to the reference traditional LNA, so the die area should not increase significantly. The presented methodology — leveraging prior designs of the traditional LNA and requiring only minor changes — also allows for implementing the solution without a significant increase in design time and effort.

\section*{Acknowledgment}
This work was supported by CAPES (Brazil), Project 176/12; CNPq; MAEC-AECID through project FORTIN (Ref. D/024124/09); FEDER program of the Junta de Andaluc\'ia, project P09-TIC-5386; and the Spanish Ministry of Economy and Competitiveness, project TEC2011-28302. The authors thank Prof. Agnes Nagy (CIME-CUJAE) for her valuable comments.

\bibliographystyle{IEEEtran}
\bibliography{references}

\end{document}